# High voltage charging system for pulsed power generators

M. Evans, B. Foy, D. Mager, R. Shapovalov and P.-A. Gourdain[1]

[1]Department of Physics and Astronomy, University of Rochester, Rochester, New York, 14627, USA

A robust and portable power supply has been developed specifically for charging linear transformer drivers, a modern incarnation of fast pulsed power generators. It is capable of generator +100 kV and -100 kV at 1 mA, while withstanding the large voltage spikes generated when the pulsed-power generator is triggered. The three-stage design combines a zero-voltage switching circuit, a step-up transformer using ferrite cores, and a dual Cockcroft-Walton voltage multiplier. The zero-voltage switching circuit drives the primary of the transformer in parallel with a capacitor. With this driver, the tank circuit naturally remain in its resonant state, allowing for maximum energy coupling between the zero-voltage switching circuit and the Cockcroft-Walton voltage multiplier across a wide range of loading conditions.

**I. Introduction:**

The High Amperage Driver for Extreme States (HADES) [1], will use 1 MA of current to heat and compress millimeter-scale material samples. HADES is composed of six linear transformer drivers (LTD) that charge in parallel and fire in series to deliver more current than a single LTD. Each LTD is a 2-meter diameter casing where 22 high voltage capacitor pairs and 22 spark gap switches[2] are arranged symmetrically around a central electrode stack. The spark gap switches are triggered accordingly to a predefined sequence to form the desired current pulse. This technology has been proposed in high voltage pulsed accelerators[3], Z-pinch drivers[4], and compression experiments.

The present paper shows a new high voltage supply design that combines a Cockcroft-Walton[5] voltage multiplier, which generates high voltage DC from a low voltage DC source, using a step-up transformer, which generates an AC voltage high enough to limit the number of stages of the voltage multiplier, and a ZVS driver[6], which generates an AC source from a DC supply. The design presented here has several advantages over existing high voltage power systems. First, the symmetric topology of the voltage multiplier generates positive and negative high DC voltages without any additional components compared to a single power supply. Second, the ZVS always drives the primary inductor of the step-up transformer at resonance. The voltage reversal on the inductor switches the gates of transistors inside the ZVS, making it a self-driven circuit. The resonant frequency was chosen to be on the order of 10 kHz, the frequency of our ferrite cores, by adding a capacitor in parallel with the inductor. The self-driven nature of the circuit allows the ZVS to track changes in the resonant frequency, caused by the transformer loading and non-linearities inherent to the ferrite cores. Finally, the input DC voltage is boosted by the resonant ZVS circuit, allowing the use of benchtop DC supplies to power the whole system. For a high frequency system like ours any VAC variations are noticeable across the transformer cores, the input would need to be rectified and regulated to avoid fluctuations. Benchtop DC power supplies already perform this function. In this manuscript, we report on the basic design, construction and characterization of this system and how it is used to charge the capacitors of a 1 GW pulsed-power system. Minor improvements, like voltage or current regulation of the power supply, have been discussed in other literature and are not reported here.

## II. Principle of operation

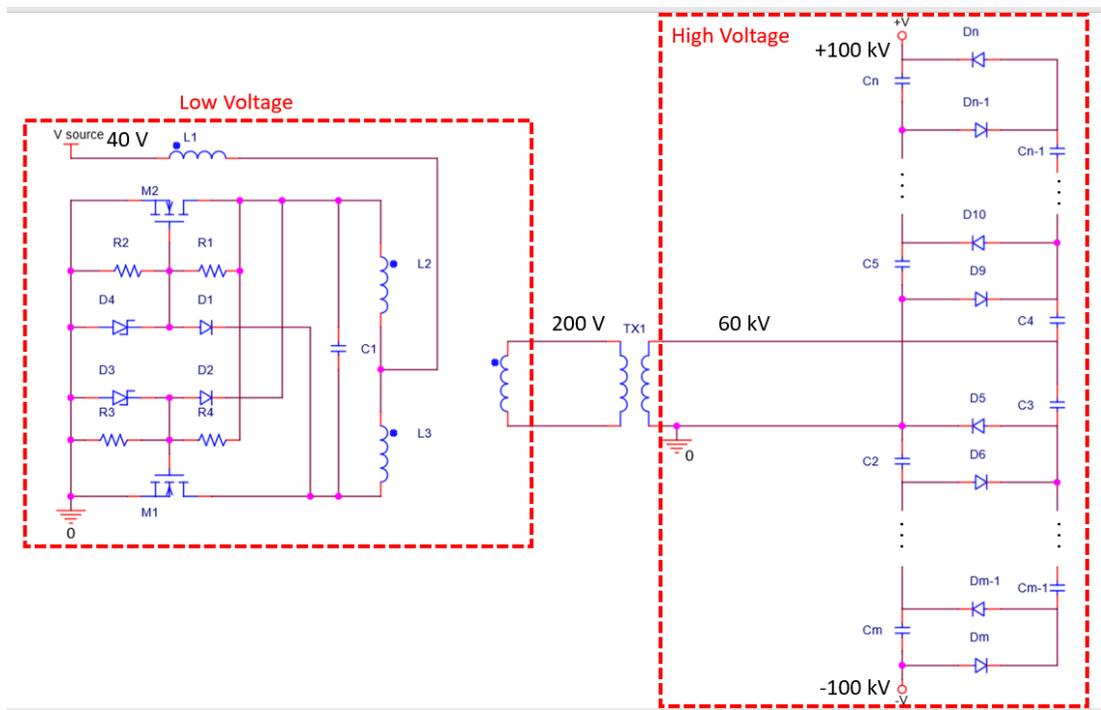

Figure 1: Circuit schematic of dual high-voltage and frequency charging system

Figure 1 shows the circuit schematic of the dual voltage power supply together with the peak voltage in the different stages of the power supply. In the first stage of the system, the ZVS driver generates a high frequency AC voltage from a low voltage DC source. The ZVS operates between 12 to 40 DC input voltage, which allows for control of the final output voltage. The ZVS outputs a 240 VAC peak-to-peak (pk-pk) at 40 Amps which is coupled to the step-up transformer. It is essential to have a high coupling factor to operate at low power. The 10-kHz transformer, with a ferrite core, has a 1:250 winding ratio between primary and secondary, raising the voltage to a maximum of 60 kV AC pk-pk. The frequency is constrained by the ferrite which operates in a 5 to 20kHz frequency range. The transformer also adds a layer of electrical insulation between the high and low voltage sides of the supply, and has a design current limit[1] of 10 mA. The AC output is then fed into the dual voltage multiplier to produce ±100 kV DC.

While coupling a transformer to a voltage multiplier is common practice in high voltage supplies, the novel idea presented in this paper is using the ZVS to always drive the transformer at resonance, under a variety of load conditions. When the primary coil is driven at resonance the magnetic field of the primary is in phase with magnetic field of the secondary winding. A strongly coupled system generates the maximum voltage on the secondary coil due to the increase of the mutual flux, reducing resistive losses on the primary coil and heat generation.[7] An efficient system will have a high coupling coefficient and less flux leakage.

### III. Topology and Spice simulation:

In the first stage, the ZVS driver converts DC into high frequency AC. A ZVS driver uses two MOSFETs that switch voltage between them with minimal loss from Ohmic heating. The schematic of a classical ZVS circuit is displayed in figure 2.

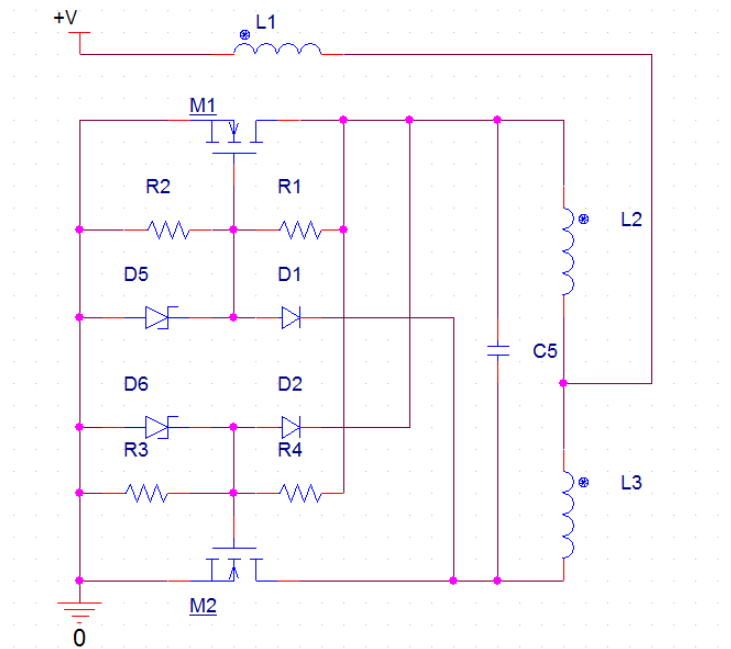

Figure 2: Schematic of ZVS driver

The principle of a ZVS works by exploiting the differences in physical electronic components of the same type and an LC oscillator[6]. When power is applied both MOSFETs see voltage on their gates and the MOSFETs begin to turn on. In an ideal circuit both MOSFETs would turn on at the exact same time and there would be no oscillation; however, because no two components are exactly alike, one MOSFET begins to turn on faster than the other. The current between gates is no longer identical and the second MOSFET begins to turn off. A capacitor couples to the transformer coil, forming an LC circuit, which then oscillates as the MOSFETs successively turn on and off.

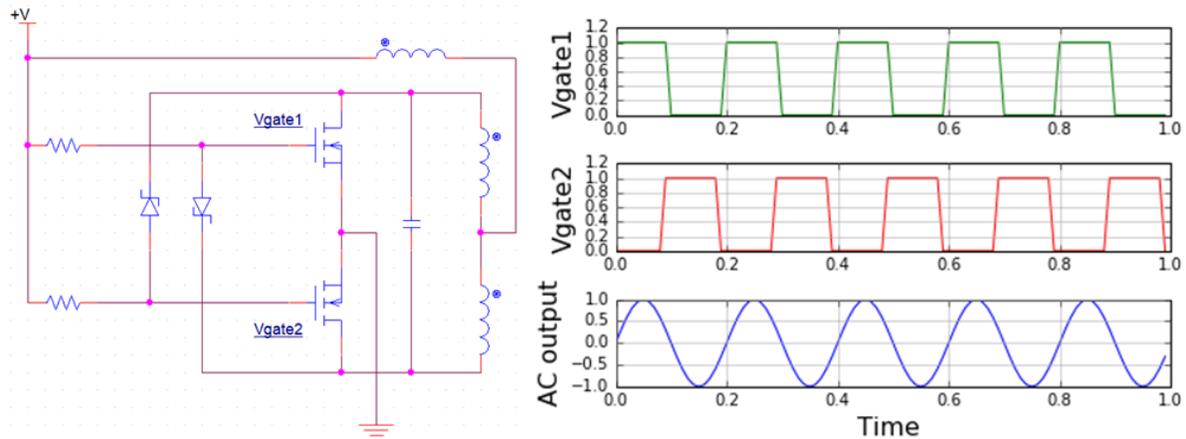

Figure 3: ZVS operating principle.

Figure 3 illustrates the voltage at each MOSFET gate and the AC output waveforms. As the voltage across MOSFET 1 begins to drop, the voltage rises on MOSFET 2. The resulting oscillation can be seen across the secondary coils as an AC voltage. The driver switches between gates exactly when the voltage across a MOSFET is zero which is why it is called zero-voltage switching. This is the time at which the MOSFET is carrying the least amount of power eliminating the need for large heat sinks.

In the second stage, the step-up transformer supplies a high voltage to the dual voltage multiplier as well as providing isolation between the high and low voltage sides. A high turn ratio is required to increase the voltage in the secondary, however to keep the high voltage transformer compact we split the secondary windings between two sides of the ferrite.

In the third stage, a dual voltage multiplier converts AC into high voltage DC via a ladder network of capacitors and diodes. Each "stage" consists of two capacitors and two corresponding diodes. The diodes rectify the

AC voltage on each side of the ladder and allow the capacitors to charge in parallel and discharge in series. The capacitors block the DC bias from the stage directly below it, allowing each stage to be at a higher potential. The voltage multiplier is a common design for generating high voltages at low currents and low costs. However, the system has a high internal impedance which makes it less attractive for high voltages, high current applications, where transformers are typically required. A schematic diagram of the dual voltage multiplier is shown in figure 4.

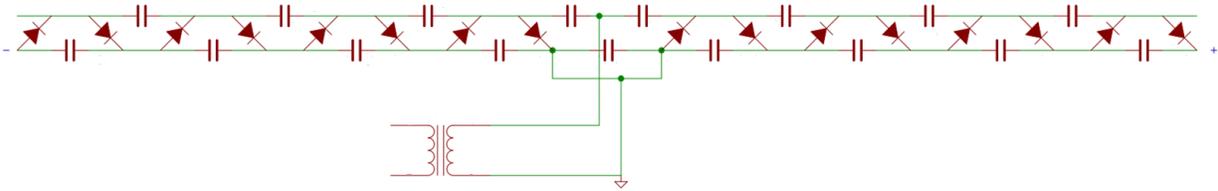

Figure 4: Schematic of dual voltage multiplier

Our design utilizes a four-stage dual voltage multiplier to produce a ±100 kV potential difference. The AC signal from the step-up transformer passes through the ladder network until it reaches ±100 kV. The higher the frequency the better the voltage amplification however the frequency is limited by the impedance of the transformer and the coupling to the ferrite circuit. The circuit along with the step-up transformer was simulated with the circuit analysis tool PSpice[8]. The ZVS circuit was replaced with a VAC signal in this simulation. The AC signal input was a sine wave of 200V pk-pk at 10 kHz. The simulation assumes ideal circuit elements. The transformer inductance was chosen based upon the ferrite material and calculated to be 10.13 µH. The output voltage with no load is illustrated below in figure 5.

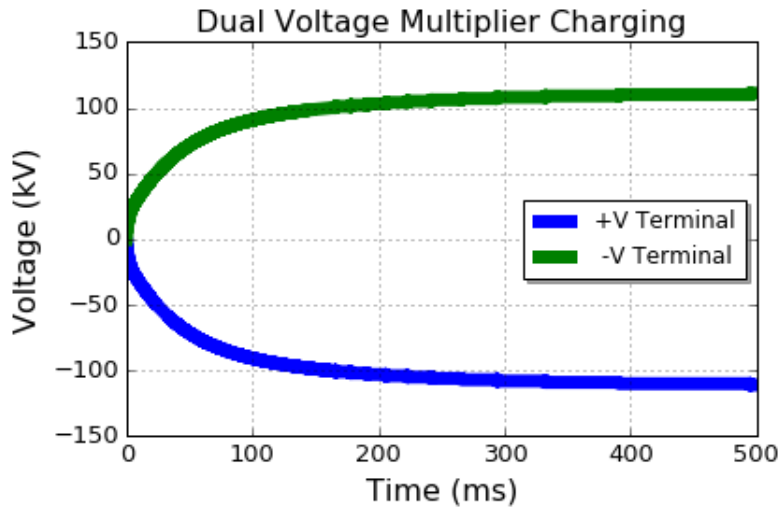

Figure 5: PSpice Simulation: Voltage vs time plot of a dual voltage multiplier charging system

The PSpice results of the voltage multiplier show the voltage slowly ramping up to max voltage. The circuit reaches a maximum of 110 kV measured at each end. The dual voltage multiplier has a charging time of roughly 300 ms.

## IV. Design Considerations and Construction:

With the exception of the ZVS, the charging system was constructed from simple components and displayed in figure 6, an extra capacitor is added to construct an interlocking pattern that reduces the distance between the two branches. The extra capacitor is grounded.

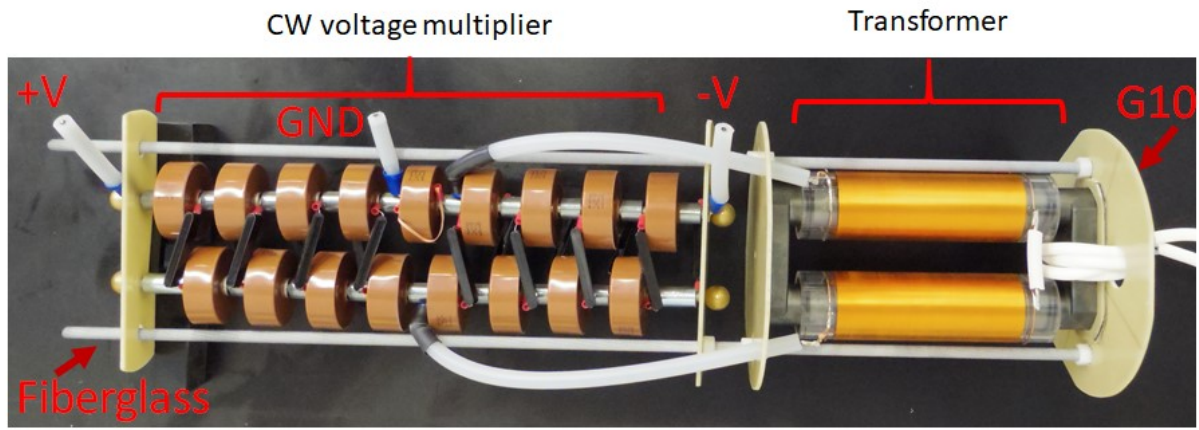

Figure 6: Assembled form. (Step-up transformer and dual voltage multiplier.)

The transformer consists of the primary and secondary windings, copper coated steel welding rod split rings, Pyrex glass tubing, the ferrite material, and Buna-N O-rings. The core shape allows for the primary and two secondary series windings on the same core. The Pyrex glass tubing, 5.08 cm diameter and 58.7 cm in length, provides the housing for the ferrite and the support for the secondary windings. Buna-N O-rings (#319) are placed between the ferrite and the Pyrex tube to prevent movement and center the ferrite inside the tube. The primary winding is a 12 AWG 60kV electrical wire (#2024 from Dielectric Sciences Inc.) at 4 turns. The secondary windings are made of two cores of 30 AWG copper magnetic wire at 500 turns connected in series to give 1000 turns. The copper coated rings are used as a mechanical stress relief and connect the secondary windings to the voltage multiplier. The MnZn UY32 ferrite[9] has an inductance factor of 5200 nH/N$^2$. It is designed to operate between 5 and 20 kHz. Our transformer design required the core to be 24.6 cm by 9.20 cm. The ferrite blocks were cut and reassembled to meet these dimensions.

The dual voltage multiplier was assembled using interconnected brass studs. To construct the positive and negative stages, we use sixteen 2700 pF (30 kV max) ceramic Z5T disk capacitors. The capacitors are complemented by sixteen diodes, (30 kV max and 200 mA), attached diagonally between their respective capacitors. The voltage multiplier is configured so the positive output is at one end and the negative output is at the opposite end, reducing the chance of arcing between output terminals. 150 kV 10 AWG insulation wire (#2121A Dielectric Sciences Inc.) is attached to each terminal output. The transformer and voltage multiplier are combined in a support structure made of G10 and two fiberglass rods. This provides the system with structural stability when placed in a PVC housing. The cylinder is then filled with STO-50 silicon transformer oil[10]. The transformer oil acts as an insulator and coolant for the ferrite core when operated in steady state. Transformer oil was chosen over epoxy for easier access to parts for maintenance. The total weight of the power supply is close to 100 lbs when filled with transformer oil.

## V. Results:

We measured the ZVS output both unloaded and then loaded with the step-up transformer and voltage multiplier. Unloaded the ZVS circuit produces a maximum of 240 VAC pk-pk signal at 15 kHz at 40 V DC input. The power supply loaded with the brick test station produces 200 VAC pk-pk at a frequency of 10 kHz.

The power supply output voltage is measured using two identical voltage dividers of 1 GΩ and 1 kΩ, into a SIGLENT SHS800 digital oscilloscope with bandwidth of 150 MHz and sampling rate of 1 GSa/s. The maximum voltage from each voltage output was ±100 kV. The current of the system is determined by creating a multi-loop circuit and applying Kirchhoff's current law. A 100 MΩ resistor is connected between the positive and negative outputs where it creates another current path. The potential difference over a 100 MΩ resistor was 200 kV, producing 2 mA of current at an output power of 400 Watts. The DC input to the ZVS was 23 V and 36 A producing 828 Watts, resulting in an efficiency of 48%.

To measure the charging time and loaded voltage we construct a brick testing station shown in figure 7, along with its circuit schematic. The brick testing station consists of 2x high voltage 47 nF capacitors, a spark gap switch, a shunt resistor, 2x 1 GΩ resistors, and 2x 250 MΩ charging resistors. One capacitor is charged to +90 kV, while the other is charged to -90 kV. The spark gap is attached between the capacitors for a total potential difference of 180 kV.

The spark gap switch was pressurized with synthetic air. The spark gap switch is self-triggered in this configuration. The brick testing station is placed in STO-50 silicon oil to prevent arcing. To measure the voltage on each capacitor a 1 GΩ voltage monitor is connected to each electrode.

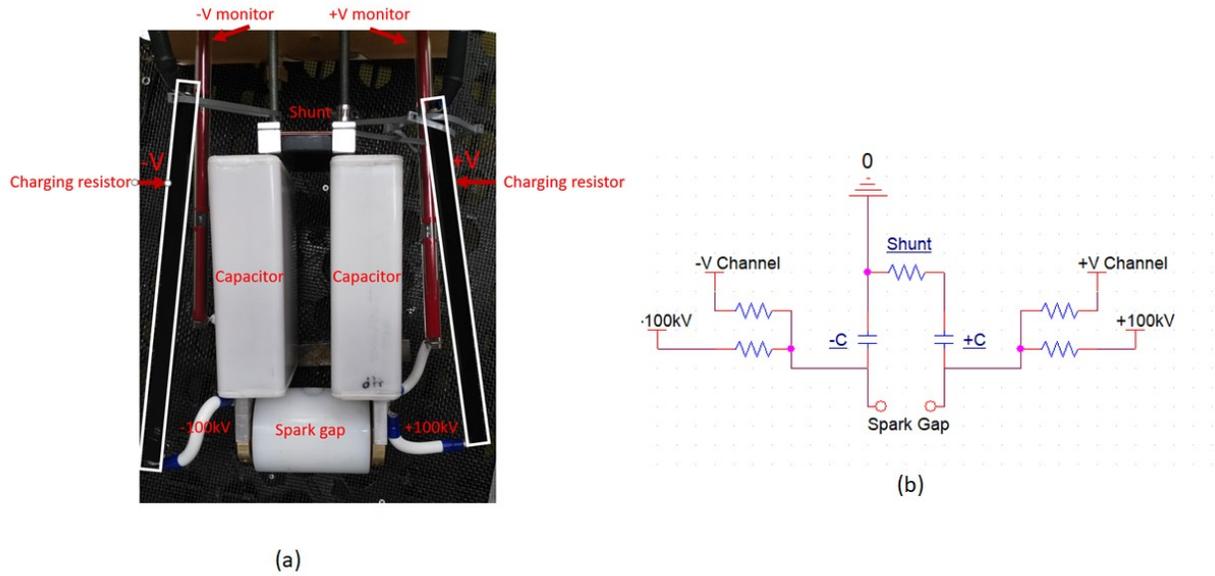

Figure 7: Brick testing station. (a) Photo of brick testing station. Charging resistors are outlined in white boxes. (b) circuit schematic

The charging resistors are made of TIVAR Cleanstat[11], a conducting plastic, with a square cross-section of 1 cm and length of 30 cm. The high resistive plastic, with resistivity 7.2 GΩ*cm, is used in the packaging industry to dissipate electrostatic charge build up on conveyor belts. Since the spark gap are self-breaking, the charging resistors are used to slow down the triggering rate of the brick by throttling the charge of the capacitor. The 1 GΩ voltage monitors are constructed by connecting four 250 MΩ resistors in series. A plot of the charging voltage across the positive load capacitor is illustrated in figure 8.

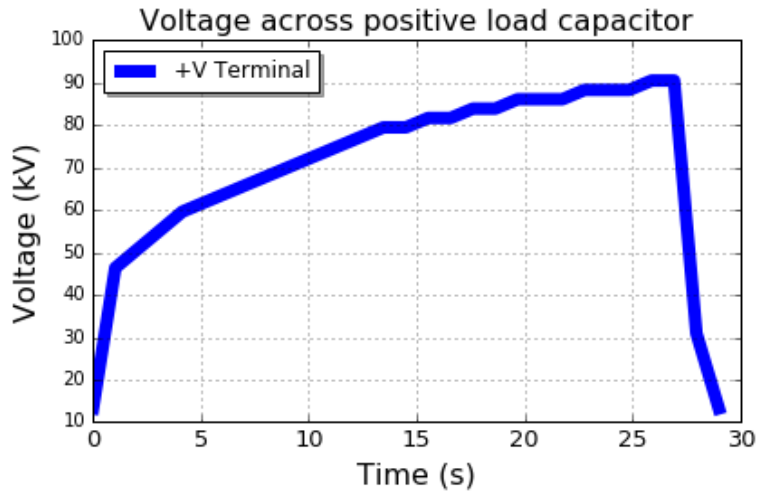

Figure 8 Load Capacitor: Voltage vs time

As seen in figure 8 the total charging time of our system was roughly 26 seconds. The max voltage at each capacitor was ±90 kV at the time of switch breakdown. The spark gap switch was self-triggered by controlling the pressure of dry air between electrodes. At ±90 kV across the capacitors the pressure needed to prevent premature breakdown is 120 psi. Once the capacitors are fully charged, breakdown occurs and generates a back electromagnetic pulse (EMP). The large back EMP caused by the spark gap would require many commercial power supplies to be disconnected before the breakdown. Or at least they would require additional protection since they are usually based on step-up transformer only. However, the proposed power supply can withstand this back EMP, since the voltage multiplier acts as a low impedance, 188Ω, short to ground for such pulses. The power supply and spark gap switch was tested in repetition for reliance and robustness over a thousand times.

## VI. Conclusion:

A dual high voltage and high frequency charging system was designed and constructed at the eXtreme State Physics Laboratory at the University of Rochester. The design is a combination of three separate systems, a ZVS circuit that supplies high frequency AC power at resonance, a step-up transformer to increase voltage and dual voltage multiplier to convert medium voltage AC to high voltage DC, while protecting the transformer from back EMP. The charging system provides ± 100 kV at 2 mA for 400 Watts of power to a 100 MΩ resistor. The self-tuning ZVS coupled with the high frequency transformer results in fast charge time for various load conditions. The power supply charges the brick test station within 26 seconds to voltage of ±90 kV before electrical breakdown. It is capable of firing thousands of shots through a spark gap switch without failure, making it a reliable system for future pulsed power experiments.

## Acknowledgments

This work is supported by the Laboratory of Laser Energetics' Horton Fellowship and the U.S. Department of Energy's National Nuclear Security Administration under Award Number DE-SC0016252.